\newcommand{\CLASSINPUTbottomtextmargin}{0.99 in}
\documentclass[journal]{IEEEtran}
\usepackage{cite}
\usepackage{amsmath,amssymb,amsfonts}
\usepackage{algorithmic}
\usepackage{graphicx}
\usepackage{textcomp}
\def\BibTeX{{\rm B\kern-.05em{\sc i\kern-.025em b}\kern-.08em
    T\kern-.1667em\lower.7ex\hbox{E}\kern-.125emX}}
\usepackage{booktabs}
\usepackage{xcolor}
\usepackage{epsfig}
\usepackage{epstopdf}
\usepackage{acronym}
\graphicspath{{fig/}}
\usepackage{mathtools}
\usepackage{array,multirow,graphicx}
\usepackage{tabularx}

\newacro{ACDD}{Alamouti with cyclic delay diversity}
\newacro{URLLC}{ultra-reliable low-latency communications}
\newacro{3GPP}{third generation partnership project}
\newacro{PHY}{physical layer}
\newacro{MIMO}{multiple-input multiple-output}
\newacro{SIMO}{single-input multiple-output}
\newacro{MISO}{multiple-input single-output}
\newacro{SISO}{single-input single-output}
\newacro{MRC}{maximum-ratio combining}
\newacro{SNR}{signal-to-noise ratio}
\newacro{CP}{cyclic prefix}
\newacro{CDD}{cyclic delay diversity}
\newacro{FSC}{frequency-selective channel}
\newacro{STC}{space-time coding}
\newacro{FFT}{fast Fourier transform}
\newacro{LMMSE}{linear minimum mean-squared error}
\newacro{FER}{frame error rate}
\newacro{OFDM}{orthogonal frequency division multiplexing}
\newacro{OCDM}{orthogonal chirp division multiplexing}
\newacro{FSC}{frequency-selective channel}
\newacro{CSI}{channel state information}
\newacro{MMSE-PIC}{mininmum mean squared error with parallel iterference cancellation}
\newacro{PFE}{perfect-feedback equalizer}
\newacro{FD}{frequency domain}
\newacro{PDP}{power delay profile}
\newacro{PDF}{probability density function}
\newacro{DFT}{discrete Fourier transform}
\newacro{ICI}{inter-carrier interference}
\newacro{OTFS}{orthogonal time frequency space}
\newacro{AWGN}{additive white Gaussian noise}
\newacro{UW}{unique word}
\newacro{ZC}{Zadoff-Chu}

\def\H{^{\rm H}}

\newcommand{\ma}[1]{\mathbf{#1}}
\newcommand{\Ht}{^{\rm H}}
\newcommand{\T}{^{\rm T}}
\newcommand{\pa}[1]{\left(#1\right)}
\def\UW{\rm UW}
\newcommand{\F}{\mathbf{F}}

\usepackage{soul,color}

\usepackage{tikz}
\usetikzlibrary{shapes,arrows}
\usetikzlibrary{positioning,calc}

\begin{document}

\title{Channel Estimation for MIMO Space Time Coded	OTFS under Doubly Selective Channels}

\author{
	\IEEEauthorblockN{Roberto Bomfin\IEEEauthorrefmark{1}, Marwa Chafii\IEEEauthorrefmark{2}, Ahmad Nimr\IEEEauthorrefmark{1} and Gerhard Fettweis\IEEEauthorrefmark{1}}
	
	\IEEEauthorblockA{\IEEEauthorrefmark{1}Vodafone Chair Mobile Communications Systems, Technische Universit{\"a}t Dresden, Germany}
	
	\IEEEauthorblockA{\small\texttt{ \{roberto.bomfin, ahmad.nimr\}@ifn.et.tu-dresden.de, gerhard.fettweis@tu-Dresden.de}}
	
	\IEEEauthorblockA{\IEEEauthorrefmark{2}ETIS, UMR8051, CY Cergy Paris Universit\'{e}, ENSEA, CNRS, France}
	
	\IEEEauthorblockA{\small\texttt{marwa.chafii@ensea.fr}} \vspace{-0.8cm}
}

\maketitle
\thispagestyle{empty}
\begin{abstract}
	In this paper, we present a unique word \mbox{(UW)-based} channel estimation approach for multiple-input multiple-output (MIMO) systems under doubly dispersive channels, which is applied to orthogonal time frequency space (OTFS) with space time coding (STC).
	The OTFS modulation has been recently proposed as a robust technique under time varying channels due to its property of spreading the data symbols over time and frequency.
	Yet another relevant aspect is the employment of multiple antennas at the transmitter and receiver. Therefore, we consider an STC MIMO system with cyclic delay diversity at the transmitter and maximum ratio combining at the receiver, where we develop a UW-based channel estimation scheme for multiple transmit antennas. 
	We show a recently proposed frame optimization scheme for SISO is directly applicable to MIMO.
	In addition, we evaluate numerically the frame error rate (FER) of OTFS and OFDM with 2$\times$2 and 4$\times$4 MIMO, where the time varying channel is estimated using the UW-based approach.
	The FER results reveal that OTFS becomes more advantageous than OFDM for MIMO-STC systems with higher order modulation and code rate.
\end{abstract}

\section{Introduction}
\IEEEPARstart{I}{t} is well known that the \ac{MIMO} technology improves the system's performance significantly by means of \ac{STC} \cite{Gore}, which is a key feature of modern communication systems.
At the receiver, the diversity is relatively straightforward to obtain. 
For instance, a common approach is the \ac{MRC} technique, which increases the \ac{SNR} after combining the signals from different receiver antennas \cite{Brennan}. 
On the other hand, obtaining diversity at the transmitter is more challenging.
In particular, schemes based on delay diversity have been introduced in~\cite{Wittneben,Winters}, which are referred to as \ac{CDD} in case of cyclic shifts, as shown in \cite{Plass,Bomfin5GWF}. 
In addition, a simple procedure was proposed by Alamouti for two transmit antennas \cite{Alamouti}.
In  \cite{Odair,Bomfin5GWF}, the authors have shown that Alamouti's \ac{STC} is directly applicable for modern systems under \acp{FSC} with frequency-domain processing, under the assumption that the channel remains static for two data block transmissions.
That is, when the channel changes among the data blocks, Alamouti's STC scheme used in \cite{Odair} introduces errors.
Since we consider a time-variant channel model in this work, the CDD transmit diversity technique is chosen.

While increasing the number of antennas is beneficial in terms of data rate and robustness, it also poses an extra challenge regarding the channel estimation, which can be critical for multiple transmit antennas.
In this work, we consider the \ac{UW} channel estimation, where a deterministic signal is sent always before the data block \cite{Coon,BomfinTWC}.
For MIMO systems, the authors in \cite{ShahabTWC} demonstrated that the UW-based channel estimation has a smaller overhead than pilot-based techniques.
When \ac{CP} is included in the data blocks, one can interpret the UW-based channel estimation as a pilot block scheme as in \cite{Zheng}, but with a variable size as \cite{Coon,ShahabTWC} such that overhead is decreased.
Recently, the authors in \cite{BomfinTWC} proposed a frame optimization tool for SISO systems, where the number of sub-blocks can be optimized for the UW based channel estimation system.
In this work, we extend the work of \cite{BomfinTWC} to MIMO.
To this end, we employ a similar scheme of \cite{ShahabTWC}, where shifted versions of UW are transmitted by each antenna.
In addition, we show that this technique is equivalent to the \ac{CDD} transmit structure, which requires a straightforward implementation.
Lastly, we prove that the frame optimization for SISO in \cite{BomfinTWC} works directly for the MIMO scheme.

Another degree of freedom to improve the performance of wireless communications systems is the waveform design.
In particular, there are several works in the literature demonstrating that the waveform design or precoding plays a significant role on this regard \cite{Bomfin,Bomfin_WCNC,AhmadOTFS}. 
For example, the authors in \cite{Bomfin} demonstrated that the well known \ac{OFDM} is suboptimal under \acp{FSC}, with the condition of \ac{CSI} being available at the receiver. 
On the contrary, the \ac{OCDM} scheme proposed in \cite{OCDM} theoretically provides optimal performance, if an iterative receiver that achieves the performance of \ac{PFE} is used. 
In \cite{Bomfin_WCNC}, an iterative receiver based on the \ac{MMSE-PIC} with low-complexity has been proposed for \ac{OCDM} with performance  approaching the \ac{PFE}.
In addition, another recently proposed waveform is \ac{OTFS} \cite{OTFS,AhmadOTFS}, which was designed for doubly dispersive channel. 
As shown in \cite{BomfinTWC}, OTFS has very similar properties as OCDM, however, OTFS also spreads the data symbols among the data sub-blocks, which is not performed in OCDM.

The main contributions of this paper are listed as follows:
\begin{itemize}
	\item Extend the SISO based channel estimation and frame optimization scheme of \cite{BomfinTWC} to MIMO. 
	\item Evaluate the performance of OTFS against OFDM under doubly dispersive channel with channel estimation. One particularly interesting outcome of this work is to show that the performance gap between OTFS and OFDM increases with high order constellation code rate.
\end{itemize}

\subsubsection*{Notations} the operator $\mathbb{E}(\cdot)$ denotes the expected value. The  matrices $\mathbf{I}_N$ and $\mathbf{F}_N$ with size $N \times N$ are  the identity and normalized Fourier matrix, respectively. The operators  $\left(\cdot\right)^\dagger$ and $\left(\cdot \right)\H$ denote conjugate and hermitian, respectively.
The circular convolution between two vectors $\mathbf{a}$ and $\mathbf{b}$  is expressed as $\mathbf{a} \circledast \mathbf{b}$ \cite[eq. (9.6.6)]{Orfanidis}, and the Kronecker product is $\mathbf{a} \otimes \mathbf{b}$. The operation ${\rm diag}(\mathbf{A})$ returns a vector corresponds to the diagonal elements of $\mathbf{A}$. $\langle n \rangle_{K}$ is the modulo operation.

\section{System model}\label{sec:system_model}
\subsection{Transmitter}
In this paper, we consider the bit-interleaved coded modulation (BICM) transmission, where a vector of information bits $\ma{b} \in \left\{0,1\right\}^{N_{\rm b}}$ of length $N_{\rm b}$ bits is encoded generating the coded bit stream $\ma{c} = {\rm enc}\pa{\ma{b}} \in \left\{0,1\right\}^{N_{\rm c}}$ of length $N_{\rm c}$. Thus, the code rate is given by  $R = {N_{\rm b}}/{N_{\rm c}}$. 
In the sequel, the coded bits $\ma{c}$ are interleaved by the operation $\ma{c}' = \Pi\pa{\ma{c}}$, where $\Pi (\cdot)$ represents the interleaver,  which is then mapped onto a QAM constellation set $\mathcal{S}$ with cardinality $|\mathcal{S}| = J$, generating the data vector $\ma{d} = {\rm mapper }\pa{\ma{c}'}\in \mathcal{S}^{N}$, with $\mathbb{E}( \ma{d} \ma{d}^{\rm H} ) = E_{\rm s}\ma{I}$, where $E_{\rm s}$ is the average energy per symbol.
In general, we consider that the transmission is split into $M$ sub-blocks of size $K = N_{\rm c}/(M \cdot \log_2 J)$, which is the amount of samples transmitted in each sub-block. 
The data vector is linearly modulated as 
%
$\ma{x} = \ma{A}\ma{d}$,
%
which can be viewed as a concatenation of $M$ sub-blocks as $\ma{x} = [\ma{x}_0\T\,\,\ma{x}_1\T\,\,\cdots\,\,\ma{x}_{M - 1}\T]\T$, for ${\ma{x}}_m = (\ma{x})_{n=mK}^{(m+1)K-1} \in \mathbb{C}^{K}$.
Moreover, $\ma{A} \in \mathbb{C}^{N \times N}$ is regarded as the linear modulation matrix.
In this work, we consider OFDM and  \ac{OTFS}, where  $\ma{A}_{\rm OFDM} = \ma{I}_M \otimes \ma{F}_K\Ht$ and $\ma{A}_{\rm OTFS} = \ma{F}_M \otimes \ma{F}_K$ \cite{AhmadOTFS}.

\subsubsection{Cyclic delay diversity}
As a transmit diversity scheme, we employ the CDD technique because it is simple and as opposed to the Alamouti STC scheme, CDD does not rely on static channels \cite{Bomfin5GWF}.
Moreover, it is not necessary to use more than one time slot for transmission. Thereby we consider only one modulated data vector $\mathbf{x}_m$. 
In general, we notice that CDD allows more than 2 transmit antennas \cite{Plass,Bomfin5GWF}.
The modulated data of CDD is given for the $n_{\rm t}$-th transmit antenna by
\begin{equation}\label{eq:x_CDD}
\ma{x}_{m}^{n_{\rm t}} = 1/\sqrt{N_{\rm t}}\boldsymbol{\delta}_{(n_{\rm t}K/N_{\rm t})} \circledast \ma{x}_m.
\end{equation}
After the CDD-STC, a \ac{CP} is inserted among the data blocks as
\begin{equation}\label{eq:x_CP}
\tilde{\ma{x}}_m^{n_{\rm t}} =
\begin{bmatrix}
(\ma{x}_m^{n_{\rm t}})_{n=N-N_{\rm CP}+1}^{N} \\ {\ma{x}}_m^{n_{\rm t}}
\end{bmatrix},
\end{equation}
where $N_{\rm CP}$ is greater than the maximum channel delay.

\subsubsection{Unique word for channel estimation}
In the UW-based channel estimation, a deterministic signal is always transmitted in the beginning and in the end of a data frame.
Usually, the \ac{ZC} sequence \cite{ZC} is employed due to its favorable property of equally spreading the energy in \ac{FD} \cite{Coon,Shahab,Zheng}.
Thus, we also utilize the \ac{ZC} as the \ac{UW} signal for channel estimation, which is given by
\begin{equation}\label{eq:ZF}
\mathbf{x}_{\rm UW}[n] = \exp(j\pi n^2/N_{\rm UW}),
\end{equation} 
for ${n \in \{0, \cdots,N_{\rm UW}-1} \}$, where $N_{{\rm UW}}$ is the UW length. 
When multiple antennas is used at the transmitter, the UW per antenna are obtained using the CDD technique analogously to equation \eqref{eq:x_CDD}, which is given by
\begin{equation}\label{eq:x_UW_nt}
\mathbf{x}_{\rm UW}^{n_{\rm t}} = 1/\sqrt{N_{\rm t}}\boldsymbol{\delta}_{(n_{\rm t}K/N_{\rm t})} \circledast \ma{x}_{\rm UW}.
\end{equation}
Then, including the CP, this signal becomes
%
\begin{equation}
\tilde{\mathbf{x}}_{\rm UW}^{n_{\rm t}} = 
\begin{bmatrix}
(\mathbf{x}_{\rm UW}^{n_{\rm t}})_{n=N_{\rm UW}-N_{{\rm CP}}+1}^{N_{\rm UW}} \\ \mathbf{x}_{\rm UW}^{n_{\rm t}}
\end{bmatrix}.
\end{equation}
Note that $\mathbf{x}_{\rm UW}$ has a length of $N_{{\rm UW}}$, that is different from the size of $\mathbf{x}_m$ in general, hence providing a higher degree of freedom for the system than the technique employed by \cite{Zheng}. 

\subsubsection{Frame}
Finally, the transmitted signal with the multiplexed data sub-blocks is given by
\begin{equation}\label{eq:x_frame}
\mathbf{x}_{\rm frame}^{n_{\rm t}}  = \left[ ({\tilde{\mathbf{x}}_{\rm UW}^{n_{\rm t}}})^{\rm T}\,\, (\tilde{\mathbf{x}}_{0}^{n_{\rm t}})^{\rm T}\,\,\cdots\,\,(\tilde{\mathbf{x}}_{M-1}^{n_{\rm t}})^{\rm T}\,\,({\tilde{\mathbf{x}}_{\rm UW}^{n_{\rm t}}})^{\rm T}\right]\T.
\end{equation}

\subsection{Wireless channel}
\subsubsection{Multiple transmit antennas}
We consider a discrete-time wireless channel, whose impulse response for the $n_{\rm t}$-th transmit antenna and $n_{\rm r}$-th receive antenna at the $n$-th time index is 
\begin{equation}\label{eq:h}
\mathbf{h}_n^{n_{\rm t},n_{\rm r}} = [h_{n,0}^{n_{\rm t},n_{\rm r}}\,\,h_{n,1}^{n_{\rm t},n_{\rm r}}\,\,\cdots\,\, h_{n,L-1}^{n_{\rm t},n_{\rm r}}]\T,
\end{equation}
where $L$ denotes the channel length, such that $h_{n,l} = 0$,  $l \notin \{0,1,\cdots,L-1\}$. Moreover, $h_{n,l}$ is stationary complex Gaussian process  w.r.t. $n$, and uncorrelated w.r.t. $l$. The average power $\rho_l = \mathbb{E}(|h_{n,l}|^2 ) $ depends on a \ac{PDP} model.
We assume the conventional channel statistics whose envelope $|\mathbf{h}_n[l]|$ follows the Rayleigh distribution for an arbitrary $n$, and phase $\Phi_{n,l} = {\rm phase}(\mathbf{h}_n[l])$ follows the uniform distribution between $0$ and $2\pi$.
In addition, we model the $l$-th path as a random process with the correlation function given by 
\begin{equation}\label{eq:E_hh}
\begin{aligned}
\mathbb{E}\left({h_{n,l}^{n_{\rm t},n_{\rm r}} {h_{n',l'}^{n_{\rm t},n_{\rm r}}}^\dagger}\right) = \left\{ \begin{matrix}
\rho_l \Upsilon(n-n')
&, l=l'
\\ 0 &, l \neq l'
\end{matrix}\right.
\end{aligned}
\end{equation}
which is independent from $({n_{\rm t},n_{\rm r}})$ and is based on the well known Jakes' model \cite{Dent}
\begin{equation}\label{eq:J0}
\Upsilon(\Delta_n) = J_0 \left(2 \pi \Delta_n  \frac{f_{\rm D} }{B} \right),
\end{equation} 
where $B$ is the bandwidth, $f_{\rm D} = f_{\rm c}v/c$ is the maximum Doppler spread which depends on the relative speed between transmitter and receiver $v$, the speed of light in vacuum $c$ and the carrier frequency $f_{\rm c}$. 
Also, \eqref{eq:E_hh} implicitly assumes no correlation among the channel taps. Lastly, the MIMO channel is considered to be spatially uncorrelated for different antennas.

\subsubsection{Equivalent single transmit antenna channel}
As we have shown in \cite{Bomfin5GWF}, the CDD scheme of \eqref{eq:x_CDD} leads to an equivalent single transmit antenna model, which is
\begin{equation}\label{eq:h_eq}
\mathbf{h}_{n}^{m,n_{\rm r}} = \sum_{n_{\rm t}=0}^{N_{\rm t}-1} \delta_{{n_{\rm t}K}/{N_{\rm t}}} \circledast \mathbf{h}_{n+\Delta_m}^{n_{\rm t},n_{\rm r}},
\end{equation}
for $n \in \left\{0,1,\cdots, K-1 \right\}$ and $m \in \left\{0,1,\cdots,M-1\right\}$, where the cyclic shift of the data in \eqref{eq:x_CDD} is transferred to the channel.
In \eqref{eq:h_eq}, the channel $\mathbf{h}_{n+\Delta_m}^{n_{\rm t},n_{\rm r}} \in \mathbb{C}^K$\footnote{For clearness of the notation, we kept the channels $\mathbf{h}_{n}^{m,n_{\rm r}}$ and $\mathbf{h}_{n+\Delta_m}^{n_{\rm t},n_{\rm r}}$ with similar notation, where the reader should be aware that the dependence of $\mathbf{h}_{n}^{m,n_{\rm r}}$ on $m$ distinguishes both variables.} has the coefficients of \eqref{eq:h} with zeros for the samples $l>L-1$.
Since the channel response varies with time in general, the time shift variable $\Delta_m = 2N_{{\rm CP}} + N_{\rm UW} + m(N_{\rm CP}+K)$ ensures that \eqref{eq:h_eq} takes only the portion of the channel in \eqref{eq:h} that is convolved with the $m$-th data sub-block. We highlight that \eqref{eq:h_eq} does not depend on $n_{\rm t}$ because this channel now is equivalent to \ac{SISO} \cite{Bomfin5GWF}.
Analogously to \eqref{eq:h_eq}, the UW channels can be written as
\begin{equation}\label{eq:h_uw_eq}
\mathbf{h}_{n}^{u,n_{\rm r}} = \sum_{n_{\rm t}=0}^{N_{\rm t}-1} \delta_{{n_{\rm t}N_{\rm WU}}/{N_{\rm t}}} \circledast \mathbf{h}_{n+\Delta_u}^{n_{\rm t},n_{\rm r}},
\end{equation}
for $n \in \left\{0,1,\cdots, N_{\rm UW}-1 \right\}$ and $u \in \left\{0,1\right\}$.
In addition, in this case $\mathbf{h}_{n-\Delta_u}^{n_{\rm t},n_{\rm r}} \in \mathbb{C}^{N_{\rm UW}}$ and the time shift is $\Delta_u = N_{{\rm CP}} + u(N_{\rm UW} + N_{{\rm CP}}+ M(K+N_{\rm CP}))$.
Finally, we note that the indexes $m$ and $u$ in \eqref{eq:h_eq} and \eqref{eq:h_uw_eq}, respectively, differentiate the sub-block channel from the UW channel.

\subsection{Receiver}
The system considered in this paper assumes perfect time and frequency synchronization for simplicity.
In the following, under a perfect synchronization condition, the discrete-time received signal for the $m$-th sub-block and $u$-th UW is modeled.
\subsubsection{Data Signals}\label{subsec:data}
For the equivalent channel of \eqref{eq:h_eq}, the received signal for the $m$-th sub-block and $n_{\rm r}$-th received antenna is
\begin{equation}\label{eq:y_m}
{\mathbf{y}}_m^{n_{\rm r}}[n] = \sum_{l=0}^{L-1}\mathbf{h}_{n}^{m,n_{\rm r}}[l] \mathbf{x}_m[\langle n-l \rangle_{K}] + \ma{w}_m^{n_{\rm r}},
\end{equation}
for $n \in \left\{0,1,\cdots, K-1 \right\}$, where the modulo operation of $\mathbf{x}_m[\langle n-l \rangle_{K}]$ is a result of the CP insertion, and $\ma{w}_m^{n_{\rm r}}$ is the \ac{AWGN} noise with power $\sigma^2$.
Since we consider an \ac{FD} equalization, it is convenient to express $\mathbf{Y}_m^{n_{\rm r}} = \F_{K}\mathbf{y}_m^{n_{\rm r}}$ in \ac{FD}, which is given by
\begin{equation}\label{eq:Y_m}
\begin{split}
\mathbf{Y}_m^{n_{\rm r}} &  \stackrel{(a)}{=} (\bar{\mathbf{\Lambda}}_m^{n_{\rm r}} + \tilde{\mathbf{\Lambda}}_{{\rm e}_m}^{n_{\rm r}})\ma{X}_m  + \ma{W}_m 
\\ &  \stackrel{(b)}{=} \hat{\mathbf{\Lambda}}_m^{n_{\rm r}}\ma{X}_m + (\tilde{\mathbf{\Lambda}}_{{\rm e}_m}^{n_{\rm r}} - \bar{\mathbf{\Lambda}}_{{\rm e}_m}^{n_{\rm r}})\ma{X}_m  + \ma{W}_m^{n_{\rm r}},
\end{split}
\end{equation}
where $(a)$ splits the FD channel matrix in two, namely, i) $\bar{\mathbf{\Lambda}}_m^{n_{\rm r}} \in \mathbb{C}^{K \times K}$ is a diagonal matrix whose coefficients are $\F_K (1/K\sum_{n=0}^{K-1}\mathbf{h}_{n}^{m,n_{\rm r}})$ taken from the averaged channel impulse response of \eqref{eq:h_eq} in FD, ii) $\tilde{\mathbf{\Lambda}}_{{\rm e}_m}\in \mathbb{C}^{K \times K}$ is an off-diagonal matrix and represents the \ac{ICI} due to Doppler spread.
Line \eqref{eq:Y_m}($b$) considers the estimated channel matrix $\hat{\mathbf{\Lambda}}_m^{n_{\rm r}} = \bar{\mathbf{\Lambda}}_m^{n_{\rm r}}+\bar{\mathbf{\Lambda}}_{{\rm e}_m}^{n_{\rm r}}$,  where $\bar{\mathbf{\Lambda}}_{{\rm e}_m}^{n_{\rm r}}$ is the channel estimation error.

\subsubsection{Maximum-ratio combining receive (MRC) diversity}\label{sec:MRC}
%
The \ac{MRC} scheme \cite{Brennan} can be applied by combining the received signals from different antennas in frequency domain in \eqref{eq:Y_m}. 
Precisely, the combination of signals according to \ac{MRC} is given by \cite{Alamouti}
\begin{equation}\label{eq:Y_eq_m}
\begin{split}
\mathbf{Y}_{{\rm eq}_m} & = \frac{\sum_{n_{\rm r} = 0}^{N_{\rm r}-1}  \hat{\mathbf{\Lambda}}_{m}^{{n_{\rm r}}\Ht} \mathbf{Y}_{m}^{n_{\rm r}}}{(\sum_{n_{\rm r} = 0}^{N_{\rm r}-1}  \hat{\mathbf{\Lambda}}_{m}^{{n_{\rm r}}\Ht} \hat{\mathbf{\Lambda}}_{m}^{{n_{\rm r}}})^{\frac{1}{2}} }
\\ & = \hat{\mathbf{\Lambda}}_{{\rm eq}_m}\mathbf{X}_m+ {\mathbf{W}}_{{\rm eq}_m},
\end{split}
\end{equation}
where ${\mathbf{W}}_{{\rm eq}_m}$ is the noise plus channel error vector. 
As a result, the \ac{SIMO} model of \eqref{eq:Y_m} collapses into a \ac{SISO} model given by the second line of \eqref{eq:Y_eq_m}, where the equivalent channel matrix is
\begin{equation}\label{eq:Lambda_SISO}
\hat{\mathbf{\Lambda}}_{{\rm eq}_m} =  \left(\sum_{n_{\rm r} = 0}^{N_{\rm r}-1}  \hat{\mathbf{\Lambda}}_{m}^{{n_{\rm r}}\Ht} \hat{\mathbf{\Lambda}}_{m}^{{n_{\rm r}}}\right)^{\frac{1}{2}}.
\end{equation}
Finally, as shown in \cite{Bomfin5GWF} it is worth noticing that \eqref{eq:Lambda_SISO} allows the employment of an equalizer based on the \ac{SISO} model without loss of generality. 
This simplification is of paramount importance to keep the receiver structure relatively simple.  

\subsubsection{UW signals}
For the UW signals, an analogous approach to \eqref{eq:y_m} and \eqref{eq:Y_m} is considered.
We provide signal in FD as
\begin{equation}\label{eq:Y_u}
\ma{Y}_{{\rm UW}_u}^{\rm n_{\rm r}}  = (\bar{\mathbf{\Lambda}}_u^{n_{\rm r}} + \tilde{\mathbf{\Lambda}}_{{\rm e}_u}^{n_{\rm r}})\ma{X}_{\rm UW}  + \ma{W}_{{\rm UW}_u}^{n_{\rm r}},
\end{equation}
where $ \bar{\mathbf{\Lambda}}_u^{n_{\rm r}} \in \mathbb{C}^{N_{\rm UW} \times N_{\rm UW}}$ is diagonal matrix ${\rm diag}(\bar{\mathbf{\Lambda}}_u^{n_{\rm r}}) = \F_{N_{\rm UW}} \bar{\mathbf{h}}_{u}^{n_{\rm r}}$, which is the averaged impulse response
\begin{equation}\label{eq:h_bar_u}
\bar{\mathbf{h}}_{u}^{n_{\rm r}} = \frac{1}{N_{\rm UW}}\sum_{n=0}^{N_{\rm UW}-1}\mathbf{h}_{n}^{u,n_{\rm r}}.
\end{equation}

\subsubsection{Iterative Receiver}
The iterative receiver of \cite{BomfinTWC} based on the \ac{MMSE-PIC} which considers the channel estimations error is used for \ac{OTFS}.
For OFDM, only one equalizer iteration is performed.

\section{UW-based MIMO Channel Estimation}\label{sec:ch_estimation}
\subsection{Least-Squares Method}
The goal of this section is to estimate the \ac{MISO} channel $\hat{\ma{\Lambda}}_m^{n_{\rm r}}$ in \eqref{eq:Y_m}.
We estimate the frequency response of the channel for both UWs using the least-square method, similarly to \cite{Zheng}.
Then, the frequency components of these estimates are interpolated in order to provide a channel estimation for the data blocks.
{Initially}, the UWs' channels in \ac{FD} $\hat{\mathbf{\Lambda}}_{{\UW}_u}^{'n_{\rm r}} \in \mathbb{C}^{N_{\rm UW}}$ are estimated as the element-wise division
\begin{equation}\label{eq:Hi_prime}
\hat{\mathbf{\Lambda}}_{{\UW}_u}^{'n_{\rm r}}  = \frac{\mathbf{Y}_{{\rm UW}_u}^{n_{\rm r}} }{\mathbf{X}_{\rm UW}},
\end{equation}
for $u=0$ and $u=1$ denoting the first and last UWs, respectively.
$\mathbf{Y}_{{\rm UW}_u} \in \mathbb{C}^{N_{\UW}}$ is the received UW signal defined in \eqref{eq:Y_u}.
Similarly, $\mathbf{X}_{\rm UW} = \mathbf{F}_{N_{\UW}}\mathbf{x}_{\rm UW} \in \mathbb{C}^{N_{\UW}} $ transmitted UW signal in FD. 
One observes that \eqref{eq:Hi_prime} estimates the compound channel in \eqref{eq:h_uw_eq} in FD.
In time domain, the channel estimation becomes
\begin{equation}\label{eq:h_u}
\hat{\ma{h}}_u^{n_{\rm r}} = \F_{N_{\rm UW}}^{\rm H}\hat{\mathbf{\Lambda}}_{{\UW}_u}^{'n_{\rm r}} = \bar{\ma{h}}_u^{n_{\rm r}}+\tilde{\ma{h}}_u^{n_{\rm r}}+\ma{w}_u^{n_{\rm r}},
\end{equation}
in which, $\bar{\ma{h}}_u^{n_{\rm r}}$ is the averaged impulse response \eqref{eq:h_bar_u}, $\tilde{\ma{h}}_u^{n_{\rm r}}$ is the time varying part of the channel and $\ma{w}_u^{n_{\rm r}}$ is the remaining AWGN noise.
By analogy to \eqref{eq:h_uw_eq}, $\hat{\ma{h}}_u^{n_{\rm r}}$ can be decomposed as
\begin{equation}\label{eq:h_hat}
\begin{split}
\hat{\ma{h}}_u^{n_{\rm r}} & \stackrel{(a)}{=} \sum_{n_{\rm t}=0}^{N_{\rm t}-1} \delta_{{n_{\rm t}N_{\rm UW}}/{N_{\rm t}}} \circledast (\bar{\ma{h}}_u^{n_{\rm t},n_{\rm r}}+\tilde{\ma{h}}_u^{n_{\rm t},n_{\rm r}}) +\ma{w}_u^{n_{\rm r}}
\\& \stackrel{(b)}{=}
\begin{bmatrix}
(\bar{\ma{h}}_u^{0,n_{\rm r}}+\tilde{\ma{h}}_u^{0,n_{\rm r}})_{l=0}^{N_{\rm UW}/{N_{\rm t}}-1} \\
\vdots \\
(\bar{\ma{h}}_u^{N_{\rm t}-1,n_{\rm r}}+\tilde{\ma{h}}_u^{N_{\rm t}-1,n_{\rm r}})_{l=0}^{N_{\rm UW}/{N_{\rm t}}-1}
\end{bmatrix} ,
\end{split}
\end{equation}
where $\bar{\mathbf{h}}_u^{n_{\rm t},n_{\rm r}} \in \mathbb{C}^{N_{\rm UW}}$ is regarded as the averaged channel impulse response for the $n_{\rm t}$-th transmit and $n_{\rm r}$-th receive antenna.
Equation \eqref{eq:h_hat}($b$) stacks the individual estimates $\hat{\mathbf{h}}_u^{n_{\rm t},n_{\rm r}}$ by taking the first $N_{\UW}/{N_{\rm t}}$ samples for each transmit antenna.
Also, we highlight that the channels of different transmit antennas do not overlap because we assume that $N_{\rm UW}/{N_{\rm t}} \geq L$.

\subsection{Channel Estimation per Transmit Antenna}
Based on \eqref{eq:h_hat}($b$), the channel estimation per transmit antenna is performed by
\begin{equation}\label{eq:Lambda_uw_u}
\hat{\mathbf{\Lambda}}_{{\UW}_u}^{n_{\rm t},n_{\rm r}} = 
\F_K\begin{bmatrix}
(\hat{\ma{h}}_u^{n_{\rm r}})_{n= n_{\rm t}N_{\UW}/{N_{\rm t}}}^{(n_{\rm t}+1)N_{\UW}/{N_{\rm t}}-1}\\
\ma{0}_{K-N_{\UW}/{N_{\rm t}}}
\end{bmatrix},
\end{equation}
where $\hat{\ma{\Lambda}}_u^{n_{\rm t},n_{\rm r}} \in \mathbb{C}^K$ has the same length as the sub-blocks.
The operation of \eqref{eq:Lambda_uw_u} can be interpreted as a time filtering and shifting on the vector $\hat{\ma{h}}_u^{n_{\rm r}}$, where the portion related to the $n_{\rm t}$-th transmit antenna is chosen.

As in \cite{Zheng}, we estimate the $k$-th frequency point of the $m$-th data block as a linear combination of the estimated channel for both UWs
\begin{equation}\label{eq:Lambda_hat}
\hat{\mathbf{\Lambda}}_m^{n_{\rm t},n_{\rm r}}[k,k] = \mathbf{C}_{m,k}\hat{\mathbf{P}}_k^{n_{\rm t},n_{\rm r}},
\end{equation}
for all $m = 0, \cdots, M-1$ and $k = 0, \cdots K-1$, where the off-diagonal elements of $\hat{\mathbf{\Lambda}}_m$ are zero in order to allow a simple one tap equalization in FD.
Also, the column vector 
\begin{equation}
\hat{\mathbf{P}}_k^{n_{\rm t},n_{\rm r}}=\left[\hat{\mathbf{\Lambda}}_{{\UW}_0}^{n_{\rm t},n_{\rm r}}[k]\,\,\hat{\mathbf{\Lambda}}_{{\UW}_1}^{n_{\rm t},n_{\rm r}}[k]\right]\T,
\end{equation}
contains the estimated channel of both UWs for the $k$-th frequency component, and the row vector $\mathbf{C}_{m,k}$ contains the interpolation coefficients, which are computed such that the error
\begin{equation}\label{eq:CE}
{\sigma}^2_{{\rm CE}_{m,k}} = \mathbb{E}\left( |\hat{\mathbf{\Lambda}}_m^{n_{\rm t},n_{\rm r}}[k,k] - {\mathbf{\Lambda}}_m^{n_{\rm t},n_{\rm r}}[k,k] |^2\right)
\end{equation}
is minimized \cite{Zheng}. 

Finally, the estimated channel for the $m$-th sub-block of the CDD transmit diversity scheme of \eqref{eq:Y_m} is obtained as

\begin{equation}\label{eq:h_hat2}
\begin{split}
\hat{\mathbf{\Lambda}}_m^{n_{\rm r}} =
\F_{K} 
\begin{bmatrix}
(\hat{\mathbf{h}}_m^{0,n_{\rm r}})_{l=0}^{N_{\rm UW}/{N_{\rm t}}-1} \\
\ma{0}_{K/N_{\rm t}-N_{\rm UW}}\\
\vdots \\
(\hat{\mathbf{h}}_m^{N_{\rm t}-1,n_{\rm r}})_{l=0}^{N_{\rm UW}/{N_{\rm t}}-1}\\
\ma{0}_{K/N_{\rm t}-N_{\rm UW}}
\end{bmatrix} ,
\end{split}
\end{equation}
where $\hat{\mathbf{h}}_m^{n_{\rm t},n_{\rm r}} = \F_{K}\Ht ({\rm diag}(\hat{\mathbf{\Lambda}}_m^{n_{\rm t},n_{\rm r}}))$ is the estimated channel in the time domain.

\subsection{Frame Optimization of \cite{BomfinTWC} to MIMO}
The authors in \cite{BomfinTWC} presented a frame optimization method for the SISO UW-based frame. In the following, we show that this method is directly applicable to the MIMO system of this paper.
Consider the channel related errors of the model in \eqref{eq:Y_m} as 
\begin{equation}\label{eq:error_eq}
	\begin{split}
		\tilde{\ma{W}}_{m}^{n_{\rm r}}  & \stackrel{(a)}{=} (\tilde{\mathbf{\Lambda}}_{{\rm e}_m}^{n_{\rm r}} - \bar{\mathbf{\Lambda}}_{{\rm e}_m}^{n_{\rm r}})\ma{X}_m 
		\\ & \stackrel{(b)}{=} \sum_{n_{\rm t}=0}^{N_{\rm t}-1}  (\tilde{\mathbf{\Lambda}}_{{\rm e}_m}^{n_{\rm t},n_{\rm r}} - \bar{\mathbf{\Lambda}}_{{\rm e}_m}^{n_{\rm t},n_{\rm r}})\mathbf{X}_m^{n_{\rm t}},
	\end{split}
\end{equation}
where line $(b)$ is obtained by combining equations \eqref{eq:h_eq}, \eqref{eq:y_m} and \eqref{eq:Y_m}, and $\mathbf{X}_m^{n_{\rm t}} = \mathbf{F}_K\ma{x}_{m}^{n_{\rm t}}$, being $\ma{x}_{m}^{n_{\rm t}}$ defined in \eqref{eq:x_CDD}.
Considering the model of \eqref{eq:Y_m} as an equivalent SISO system, the authors of \cite{BomfinTWC} have shown that the channel related errors can be split into two components, namely, channel estimation error and Doppler spread error. In particular, this quantities are given respectively as ${\sigma}^2_{{\rm CE}_{m,k}}$ in equation $\eqref{eq:CE}$, and $\sigma^2_{{\rm DE}_{m,k}} = \left(\mathbb{E}\left(\tilde{\mathbf{\Lambda}}_{{\rm e}_m}^{n_{\rm r}}  {\tilde{{\ma{\Lambda}}}_{{\rm e}_m}^{{n_{\rm r}}\Ht}} \right)\right)_{k,k}$. Thus, as in \cite{BomfinTWC}, the overall channel error power of \eqref{eq:error_eq} is
\begin{equation}\label{eq:error}
	\left(\mathbb{E}\left(\tilde{\ma{W}}_{m}^{n_{\rm r}} {\tilde{\ma{W}}_{m}^{{n_{\rm r}}\Ht}}\right)\right)_{k,k} = {\sigma}^2_{{\rm CE}_{m,k}}+ \sigma^2_{{\rm DE}_{m,k}}.
\end{equation}
Then, due to the independence of channels of different transmit antennas, we have $\mathbb{E}\left({\tilde{\ma{\Lambda}}}_{{\rm e}_m}^{n_{\rm t},n_{\rm r}} {\tilde{{\ma{\Lambda}}}_{{\rm e}_m}^{{n'_{\rm t},n_{\rm r}}\Ht}}\right) = \mathbb{E}\left({\bar{\ma{\Lambda}}}_{{\rm e}_m}^{n_{\rm t},n_{\rm r}} {\bar{{\ma{\Lambda}}}_{{\rm e}_m}^{{n'_{\rm t},n_{\rm r}}\Ht}}\right) = \ma{0}_{K,K}$ for $n_{\rm t} \neq n_{\rm t}'$. And due to normalization of \eqref{eq:x_CDD}, we have $\mathbb{E} (\mathbf{X}_m^{n_{\rm t}} {\mathbf{X}_m^{n_{\rm t}}} \Ht) = 1/N_{\rm t}\ma{I}_K$.
Now, combining the two observations above with line $(b)$ of \eqref{eq:error_eq}, one can verify that the channel errors per transmit antenna is equal to $({\sigma}^2_{{\rm CE}_{m,k}}+ \sigma^2_{{\rm DE}_{m,k}})/N_{\rm t}$. In other words, SISO system has the same channel errors as the multiple transmit antenna system because making $N_{\rm t} = 1$ leads to the same quantity of \eqref{eq:error}.

For multiple receive antennas with the \ac{MRC} model of \eqref{eq:Y_eq_m}, it is shown in  \cite{Bomfin5GWF} the noise term $\tilde{\mathbf{W}}_{{\rm eq}_m}$ has the same covariance matrix as the respective quantity for an arbitrary receive antenna, because the noise and channel error is independent for different receive antennas. 
Basically, it means that the channel errors of \eqref{eq:Y_eq_m} is the same as equation \eqref{eq:error}.

As a conclusion, the frame optimization framework of \cite{BomfinTWC} initially conceived for a SISO system can be straightforwardly applied to the MIMO channel estimation scheme of this paper, where the model for a single antenna is considered.

\section{Numerical Results}\label{sec:numerical_results}

In this section, the MIMO channel estimation algorithm presented in Sec.~\ref{sec:ch_estimation} is applied to OTFS and OFDM  with MIMO STC presented in Sec.~\ref{sec:system_model}.
The channel parameters are taken from \cite{BomfinTWC} for $N=288$. 
In particular, the wireless channel follows Extended Vehicular-A (EVA) model used in \cite{BomfinTWC}. 
The bandwidth is equal to $B=4.32$ MHz, leading to a channel with $L=12$ taps based on the EVA PDP. The assumed carrier frequency is $f_{\rm c} = 5.9$ GHz and the mobility condition considers a relative speed between transmitter and receiver equal to $v = 350$ km/hr, resulting in a maximum Doppler shift of $f_{\rm D} = 1.92$ kHz.

\subsection{Channel Estimation Error}
\begin{figure}[t!]
	\includegraphics[scale=0.75]{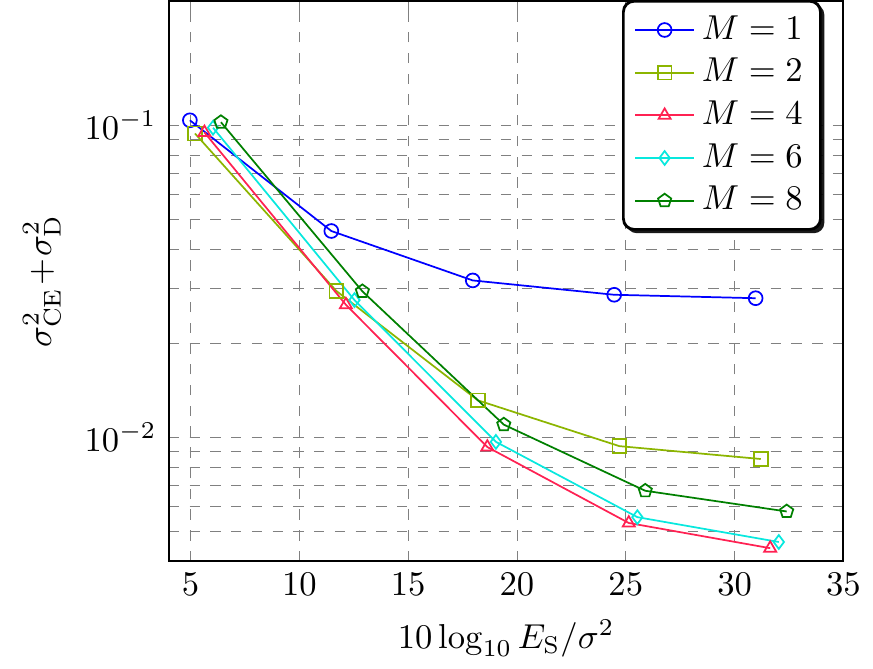}		
	\caption{Channel related errors for different $M$.}
	\label{fig:CH} 
	\vspace{-0.3cm}
\end{figure}

In Figure~\ref{fig:CH}, we show the results of the channel related errors for different values of $M=\left\{1,2,4,6,8\right\}$ such that $M K = 288$ remains constant.
In this case, we considered the UW length of $N_{\rm UW} = 32$ and a 2$\times$2 MIMO configuration. This guarantees the limit  $N_{\rm t} \leq \lfloor N_{\rm UW}/L \rfloor = 2$ in order to guarantee an independent channel estimation per transmit antenna. The CP length is $N_{\rm CP} = 16$, which is greater than $L$.
As done in \cite{BomfinTWC}, we consider the averaged channel estimation error over all sub-carriers and sub-blocks as ${\sigma}^2_{{\rm CE}} = 1/(M\cdot K)\sum_{m=0}^{M-1}\sum_{k=0}^{K-1}{\sigma}^2_{{\rm CE}_{m,k}}$. An equivalent approach is considered for the Doppler spread ${\sigma}^2_{{\rm D}} = 1/(M\cdot K)\sum_{m=0}^{M-1}\sum_{k=0}^{K-1}{\sigma}^2_{{\rm D}_{m,k}}$.
As it has been discussed in \cite{BomfinTWC}, there is a trade-off between on ${\sigma}^2_{{\rm CE}}$ and ${\sigma}^2_{{\rm D}}$ for different $M$. Basically, if $M$ increases, the channel estimation error ${\sigma}^2_{{\rm CE}}$ augments because of more CPs are included inside the data blocks. This effect makes the UW to be further away in time, which impacts negatively in the channel estimation quality. Conversely, increasing $M$ imlies in decreasing $K$, meaning that the channel is more static within one sub-block. This effect decreases the Doppler spread, and therefore decreases ${\sigma}^2_{{\rm D}}$. 
For this channel configuration, we observe that $M=4$ provides the best trade-off, which is used in the performance evaluation in the following.

\subsection{Performance}
\begin{figure}[t!]
	\includegraphics[scale=0.75]{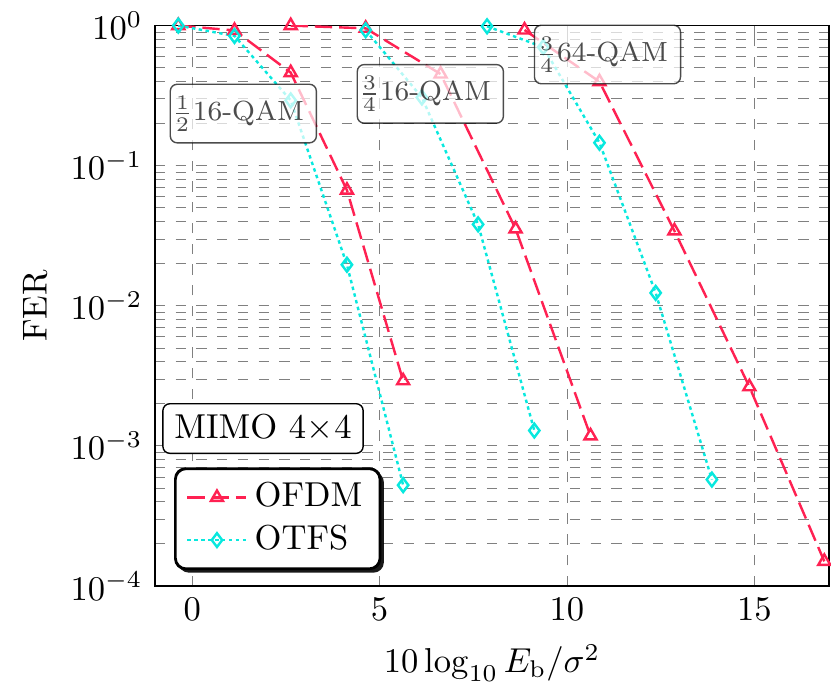}	
	\includegraphics[scale=0.75]{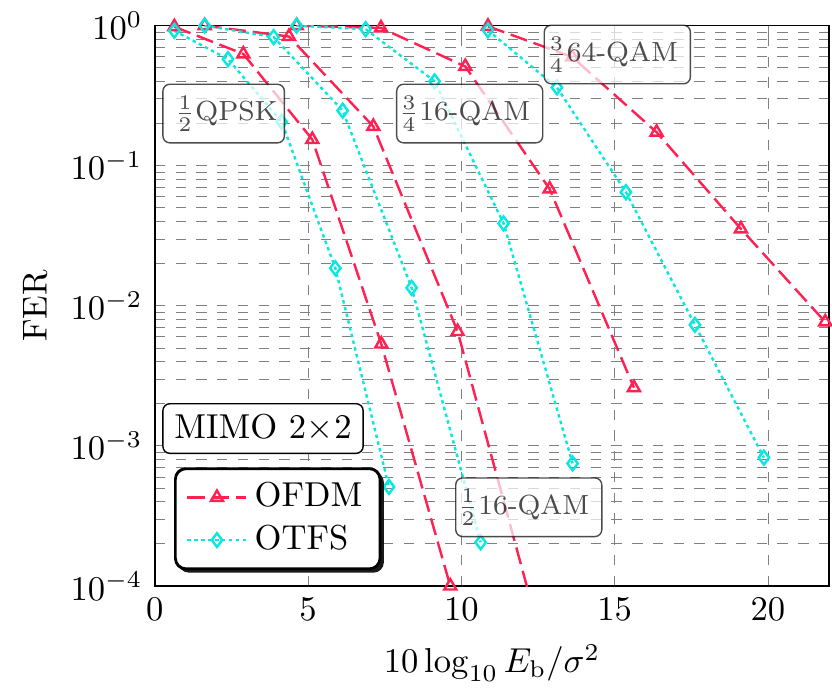}	
	\caption{FER comparison of OFDM and OTFS with SISO and MIMO.}
	\label{fig:FER} \vspace{-0.6cm}
\end{figure}

For the performance curves, we consider $M=4$ and $K=72$.
Also, the UW length of $N_{\rm UW} = 32$ is used for the 2$\times$2 MIMO configuration.
For the 4$\times$4 MIMO, $N_{\rm UW} = 64$ is chosen to attain the condition $N_{\rm t} \leq \lfloor N_{\rm UW}/L \rfloor = 5$.
For channel coding,  the $\left\{133,171\right\}_8$ recursive systematic convolutional (RSC) encoder is employed with code rates $R=\frac{1}{2}$ and $\frac{3}{4}$. The modulation coding scheme (MCS) are  $\left\{\frac{1}{2}{\rm QPSK}, \frac{1}{2}{\rm 16\text{-QAM}},\frac{3}{4}{\rm 16\text{-QAM}},\frac{3}{4}{\rm 64\text{-QAM}} \right\}$. 
The code rate $R=\frac{3}{4}$ is obtained by puncturing $66 \%$ of the parity bits generated by the half code rate RSC encoder 
The iterative receiver of OTFS is based on the LMMSE-PIC of \cite{BomfinTWC}.
The outcomes are shown in Fig.~\ref{fig:FER}.

The 2$\times$2 MIMO is shown in the top graph of Fig.~\ref{fig:FER}.
The FER is plotted against $10 \log_{10 }E_{\rm b}/\sigma^2$ per receive antenna, such that a fair comparison is made with the SISO system.
It is interesting to note that for the system with higher code rate and constellation order, namely, $\frac{3}{4}{\rm 16QAM}$, the performance gap of OTFS and OFDM is increased.
This result is explained based on the coded modulation capacity curves of \cite{Bomfin}, where in the regions of higher code rate, the information rate gap between OCDM and OFDM is increased, which is also an expected behavior with OTFS because OCDM is also a spreading waveform.
In particular, this gap is larger than 5 dB, which represents a significant gain.
The OTFS system with $\frac{3}{4}{16\text{-QAM}}$ has also a considerable performance gain in relation to OTFS of more than 2 dB.
This is a very interesting outcome, because if one considers only the curves with $R=\frac{1}{2}$ as the work in \cite{Bomfin5GWF}, it might lead to an erroneous conclusion that employing spreading waveforms such as OTFS with many RX antennas brings very small gain.
On the contrary, this outcome reveals that for higher MCS, we still obtain a non negligible gap even when the amount of receive antennas is increased.
In order to verify this behavior with more receive antennas, the 4$\times$4 MIMO system is investigated in the bottom graph.
As expected, the performance gap is decreased. However, the system with $\frac{3}{4}{64\text{-QAM}}$ still provides a performance gain of approximately 2 dB at a FER of $10^{-3}$. 

These experiments reveal a trend where OTFS systems with higher constellation order and code rate provide a greater performance improvement in relation to OFDM.
As a consequence, this enhancement is non negligible even with several receive antennas, where the performance difference tends to vanish.

\vspace{-0.1cm}
\section{Conclusion}\label{sec:conclusion}
In this paper, we have developed a unique word (UW)-based MIMO channel estimation algorithm for doubly dispersive channels which has been applied to the OTFS modulation with space time coding.
Basically, the channel estimation approach consists of generating the UW for each transmit antenna using the cyclic delay diversity (CDD) technique.
We have shown that the channel of each transmit antenna can be estimated independently if $N_{\rm UW}/N_{\rm t} \geq L$, being $N_{\rm UW}$, $N_{\rm t}$ and $L$ the UW size, number of transmit antennas and channel length in samples, respectively.
This approach is advantageous because it is simple and uses CDD technique, which can be also used as STC for the data. 
Moreover, we have demonstrated that a recently proposed frame optimization scheme developed for SISO is directly applicable for MIMO.

We have numerically compared OTFS with OFDM in terms of frame error rate (FER).
The most relevant outcome is that higher MCS increases the performance gap in favor of OTFS against OFDM.
Interestingly, the system with $\frac{3}{4}$64-QAM presents a non negligible performance gap of 2 dB even when 4 antennas are employed at the receiver, which does not happen for the systems with low code rate, e.g., $\frac{1}{2}$16-QAM.

In summary, with the support of the MIMO channel estimation scheme presented, the FER results revealed that OTFS becomes more advantageous in relation to OFDM for MIMO-STC systems with higher order QAM and code rate, which is the case for high throughput systems.
\vspace{-0.2cm}
\section*{Acknowledgments}
\small
This project has received funding from the European Union’s Horizon 2020 research and innovation programme through the project iNGENIOUS under grant agreement No 957216.
And the authors acknowledge CY Paris Initiative for the support of the project through the ASIA Chair of Excellence Grant (PIA/ANR-16-IDEX-0008).
\vspace{-0.2cm}
\bibliographystyle{ieeetr}
\bibliography{references_ha}{}

\end{document}